\input{aipcheck}

\documentclass[
    ,final            
  ]
  {aipproc}

\layoutstyle{6x9}

\begin{document}

\title{Supersymmetric Lepton Flavor Violation}


\author{Amon Ilakovac}{
  address={University of Zagreb, Department of Physics, Bijeni\v cka
cesta 32, P.O.Box 331, Zagreb, Croatia} }

\author{Apostolos Pilaftsis}{
  address={School of Physics and Astronomy, Univ.~of Manchester, 
Manchester M13 9PL, United Kingdom}
}

\begin{abstract}
 We study a new supersymmetric mechanism for lepton flavor violation
 in a minimal extension of the MSSM with low-mass heavy singlet
 neutrinos, which is fully independent of the flavour structure of the
 soft SUSY breaking sector.  We find that $\ell \to \ell' \gamma$
 processes are forbidden in the SUSY limit, whilst the processes
 $\ell \to \ell'\ell_1\ell_2$ and $\mu\to e$ conversion in nuclei can
 be enhanced well above the observable level, via large
 neutrino Yukawa-coupling effects.
\end{abstract}

\maketitle


We  present  a new  mechanism  of  lepton  flavor violation  (LFV)  in
supersymmetric  theories~\cite{PRL_IP}.  The mechanism  is independent
of the flavour  structure of soft SUSY breaking  sector and its origin
is   fully   supersymmetric.  Therefore,   we   call  this   mechanism
supersymmetric LFV (SLFV).  To illustrate  the details of SLFV, let us
assume an $R$-parity conserving seesaw  extension of the MSSM with one
singlet heavy neutrino per  generation (MSSM3N).  The leptonic part of
the MSSM3N is given by {\small
\begin{equation}
  \label{Wpot}
W_{\rm lepton}\; =\; \widehat{E}^C {\bf h}_e \widehat{H}_d
\widehat{L}\: +\: \widehat{N}^C {\bf h}_\nu \widehat{L} \widehat{H}_u\:
+\: \widehat{N}^C {\bf m}_M \widehat{N}^C\; .
\end{equation}} 
Here $\widehat{H}_{u,d}$, $\widehat{L}$, $\widehat{E}$ and
$\widehat{N}^C$ denote the two Higgs-doublet superfields, the three
left- and right-handed charged-lepton superfields and the three
right-handed neutrino superfields, respectively. The Yukawa matrices,
${\bf h}_{\nu,e}$ and Majorana mass matrix ${\bf m}_M$ are complex
3-by-3 matrices.  In a minimal supergravity (mSUGRA) approach to 
MSSM3N, the soft SUSY
breaking usually satisfies universal conditions at GUT scale.
Moreover, singlet neutrino masses are two to four orders of magnitude 
below the GUT scale to account for the observable light neutrinos via
the usual sessaw mechanism. 
In this case, the heavy neutrino LFV contributions are suppressed by a
factor $m_\nu/M_N$, with $m_{\nu} \stackrel{<}{{}_\sim}0.1$~eV
~\cite{CL}, and
LFV can  induced only sizeably through radiatively induced off-diagonal SUSY
breaking parameters, such as ~${\bf \widetilde{M}}^2_{L,E}$ and 
${\bf A}_e$~\cite{BM,HMTY}.  This soft-SUSY breaking mechanism 
represents the standard paradigm for LFV in SUSY models.

In stark contrast to soft LFV~\cite{BM,HMTY}, in supersymmetric models
with low-scale singlet neutrinos, a different source of LFV can become
dominant   which  originates   from  large   neutrino  Yukawa-coupling
effects~\cite{PRL_IP}.   This can  naturally take  place  in low-scale
seesaw models~\cite{WW,MV,BGL,AZPC}, where  the smallness of the light
neutrino  masses  is  accounted  for  by  quantum-mechanically  stable
cancellations~\cite{AZPC}  due to the  presence of  approximate lepton
flavor  symmetries~\cite{AZPC,APRLtau},   implying  the  existence  of
nearly  degenerate heavy  Majorana neutrinos  (${\bf m}_M  \approx m_N
{\bf 1}$).   These approximate  flavour symmetries allow  the Majorana
mass  scale $m_N$  to be  as  low as  100~GeV.  In  these models,  LFV
transitions  from  a  charged  lepton  $l=e,\,\mu,\,\tau$  to  another
$l'\neq l$ are enhanced by the ratios~\cite{KPS,IP,DV} {\small
\begin{equation}
  \label{Omega}
\mbox{\boldmath{$\Omega$}}_{\ell\ell'} \; =\;  \frac{v^2_u}{2  m^2_N}\ ({\bf  h}^\dagger_\nu
{\bf h}_\nu)_{ll'}
\end{equation}
}
and  are not  constrained by  the usual  seesaw  relation $m_\nu/m_N$,
where   $v_u/\sqrt{2}  \equiv  \langle   H_u\rangle$  is   the  vacuum
expectation value  (VEV) of the  Higgs doublet $H_u$,  with $\tan\beta
\equiv  \langle  H_u\rangle/\langle  H_d\rangle$. 

To assess the significance of SLFV, we  assume that the soft
SUSY-breaking scale $M_{SUSY} \ll m_N$,  
the superpotential $\widehat{H}_u\widehat{H}_d$-mixing parameter
$\mu \ll m_N$, and that ${\bf  \widetilde{M}}^2_{L,E}$   
and  ${\bf  A}_e$  are  flavor conserving, e.g.~proportional to {\bf 1} 
at the energy-scale $m_N$.

\begin{figure}[t]
 \centering
  \includegraphics[clip,width=1.00\textwidth,height=0.095\textheight,
     angle=0]{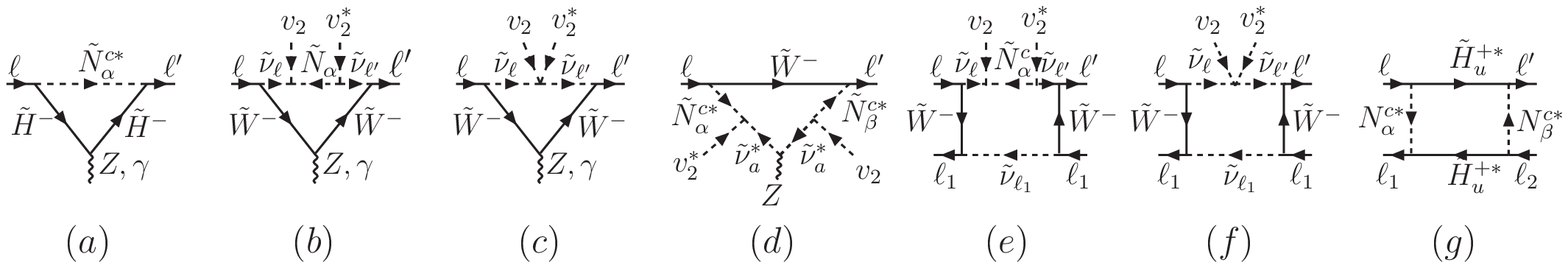}
\caption{Feynman  graphs giving rise  to leading  SLFV effects  in the
lowest   order  of   an  expansion   in  $v_u$  and
$m^{-1}_N$. Not  shown are diagrams  obtained by replacing  the tilted
SUSY        states       $\widetilde{H}^-_u$,       $\widetilde{W}^-$,
$\widetilde{N}_{\alpha}$  and   $\tilde{\nu}_l$  with  their  untilted
counterparts.}\label{f1}
\end{figure}

Within the above framework, we calculate the leading SLFV amplitudes 
close to the SUSY limit
in the lowest order of $v_u$ and $m_N^{-1}$. 
The leading order 
diagrams in $g_W$ and ${\bf h}_{\nu}$ are given in Fig. 1. 
In a self-explanatory notation, the pertinent 
LFV amplitudes read
{\small 
\begin{eqnarray}
  \label{Trans}
{\cal T}^{\gamma l'l}_\mu \!\!&=&\!\! \frac{e\, \alpha_w}{8\pi M^2_W}\
\bar{l}' \Big( F_\gamma^{l'l}\, q^2 \gamma_\mu P_L + G^{l'l}_\gamma\,
i\sigma_{\mu\nu} q^\nu m_l P_R \Big) l\;,\nonumber\\
{\cal T}^{Z l'l}_\mu \!\!&=&\!\!  \frac{g_w\, \alpha_w}{8\pi \cos\theta_w}\
F^{l'l}_Z\, \bar{l}' \gamma_\mu P_L l\; ,\\
{\cal T}^{l'l_1l_2}_l \!\!\!&=&\!\!  -\frac{\alpha^2_w}{4 M^2_W}\;
F^{ll'l_1l_2}_{\rm box}\, \bar{l}'\gamma_\mu P_L l\;
\bar{l}_1\gamma^\mu P_L l_2 \; ,\nonumber
\end{eqnarray}} 
and $q = p_{\ell'} - p_\ell$. The amplitudes {\small ${\cal
T}^{l'u_1u_2}_l$} and {\small ${\cal T}^{l'd_1d_2}_l$} have the same
structure as amplitude {\small ${\cal T}^{l'l_1l_2}_l$} up to
replacements $\ell_i \to u_i \to d_i,\ i=1,2$.  The form factors
{\small $F^{l'l}_\gamma$}, {\small $G^{l'l}_\gamma$}, {\small
$F^{l'l}_Z$}, {\small $F^{ll'l_1l_2}_{\rm box}$}, {\small
$F^{ll'u_1u_2}_{\rm box}$} and {\small $F^{ll'd_1d_2}_{\rm box}$}
receive contributions from both the heavy neutrinos $N_{1,2,3}$ and
the right-handed sneutrinos $\widetilde{N}_{1,2,3}$. To illustrate
SLFV effects we give explicit form of the form factors {\small
$F^{l'l}_\gamma$}, $G^{l'l}_\gamma$ and $F^{l'l}_Z$ in the Feynman
gauge, {\small%
\begin{eqnarray}
  \label{Fgamma}
(F^{l'l}_\gamma)^N 
 \!\!\!\!\!&=&\!\!\!\!\! 
 \frac{ \mbox{\boldmath{$\Omega$}}_{\ell\ell'} }{6\,s^2_\beta}\,
  \ln \frac{m^2_N}{M^2_W}\; ,
\qquad
(F^{l'l}_\gamma)^{\widetilde{N}} 
 \, =\, 
  \frac{\mbox{\boldmath{$\Omega$}}_{\ell\ell'}}{3\,s^2_\beta}\,
  \ln \frac{m^2_N}{\widetilde{m}^2_h}\; ,
\\
  \label{Ggamma}
(G^{l'l}_\gamma)^N 
 \!\!\!\!\!&=&\!\!\!\!\! -\,
  \mbox{\boldmath{$\Omega$}}_{\ell\ell'}\, 
  \bigg(\, \frac{1}{3\,s^2_\beta} + \frac{1}{6}\, \bigg)\;,
\qquad
(G^{l'l}_\gamma)^{\widetilde{N}} \, = \,
\mbox{\boldmath{$\Omega$}}_{\ell\ell'}\, \Bigg(\, \frac{1}{3\,s^2_\beta}  +
  \frac{M^2_W}{6\,\widetilde{m}^2_1}\,\bigg)
\; ,
 \\
  \label{FZ}
(F^{l'l}_Z)^N 
 \!\!\!\!\!&=&\!\!\!\!\! 
 -\, \frac{3\, \mbox{\boldmath{$\Omega$}}_{\ell\ell'}}{2}\,
  \ln \frac{m^2_N}{M^2_W} - \frac{( \mbox{\boldmath{$\Omega$}}_{\ell\ell'}^2) }{2\,s^2_\beta} \,
  \frac{m^2_N}{M^2_W}\; , \quad
(F^{l'l}_Z)^{\widetilde{N}} 
 \!\, =\, \! 
 \frac{\mbox{\boldmath{$\Omega$}}_{\ell\ell'}}{2} \,
  \ln \frac{m^2_N}{\widetilde{m}^2_1} + \frac{(
  \mbox{\boldmath{$\Omega$}}_{\ell\ell'}^2)_{l'l}}{4\,s^2_\beta} \,
  \frac{m^2_N}{M^2_W}\; \ln\frac{m^2_N}{\widetilde{m}^2_2}\; .\quad
\end{eqnarray}}
The   form  factors   {\small   $F^{ll'l_1l_2}_{\rm  box}$},   {\small
$F^{ll'u_1u_2}_{\rm box}$} and {\small $F^{ll'd_1d_2}_{\rm box}$}, and
masses        $\widetilde{m}^2_h$,       $\widetilde{m}^2_1$       and
$\widetilde{m}^2_2$ are given in~\cite{PRL_IP}.  Note that in the SUSY
limit      $\tan\beta      \to       1$,      $\mu\to      0$      and
$\widetilde{m}^2_h, \widetilde{m}^2_1, \widetilde{m}^2_2\ \to\ M_W^2$.

Notice that the photonic dipole form factor {\small $G^{l'l}_\gamma =
(G^{l'l}_\gamma)^N + (G^{l'l}_\gamma)^{\widetilde{N}}$} vanishes in
the SUSY limit, while beyond the SUSY limit it strongly depends on the
SUSY breaking sector. That is a consequence of the SUSY
non-renormalization theorem~\cite{FR}.  

In all other form factors, but $F^{l'l}_Z$, $N$ and $\widetilde{N}$
contributions add constructively.  Although in $F^{l'l}_Z$ $N$ and
$\widetilde{N}$ contributions add destructively, $F^{l'l}_Z$ is
strongly enhanced in the large $m_N$ limit by the logarithmic factor
in $\widetilde{N}$ contribution. The $m_N$ limit corresponds to the
large neutrino Yukawa couplings ${\bf h}_\nu$ (see Eq. (\ref{Omega})),
and in that limit the $\mbox{\boldmath{$\Omega$}}^2$ terms dominate in
$Z$ and leptonic box amplitudes.

\begin{figure}[t]
 \centering
 \includegraphics[clip,width=0.45\textwidth,height=0.20\textheight]{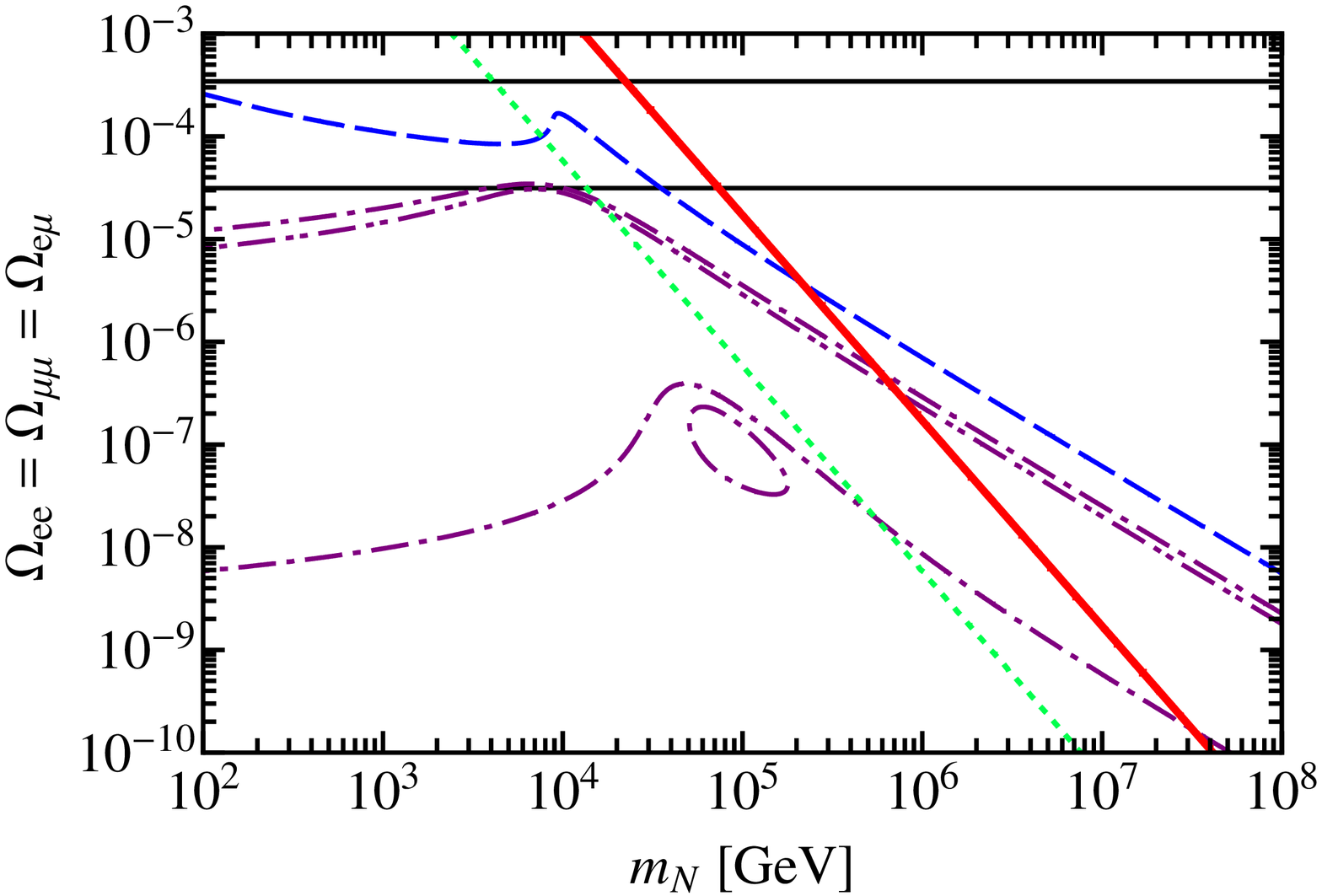}\qquad
 \includegraphics[clip,width=0.45\textwidth,height=0.20\textheight]{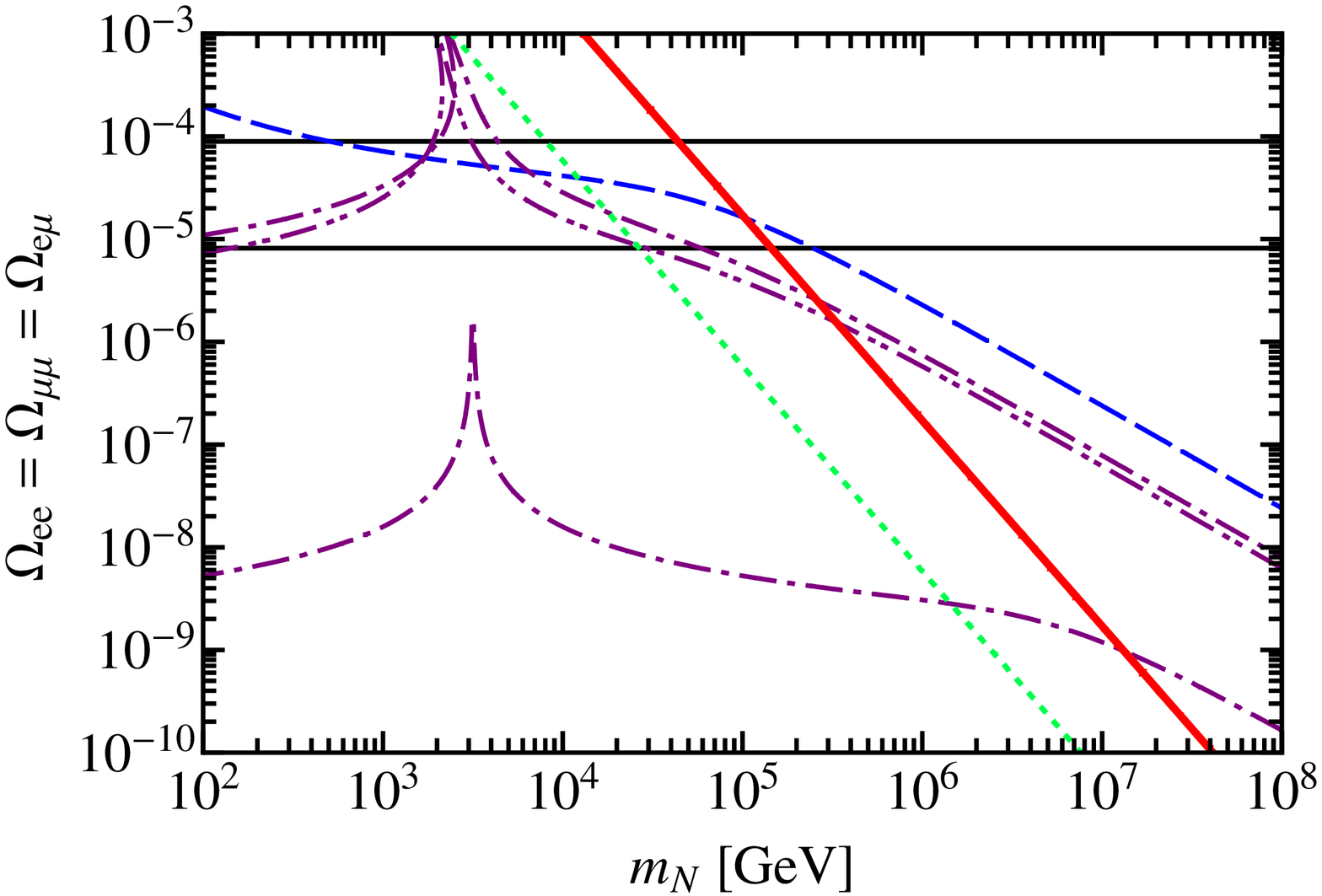}
\caption{Exclusion  contours  of  $\mbox{\boldmath{$\Omega$}}_{e \mu}$  
  versus  $m_N$ for $\mbox{\boldmath{$\Omega$}}_{\ell\ell'}$ values defined in
  the text.
  The contour lines are defined by experimental limits
  and future sensitivities: $B(\mu^- \to e^-\gamma)
  < 1.2\times 10^{-11}$~\cite{PDG} (upper horizontal line), $B(\mu^- \to
  e^-\gamma) \sim 10^{-13}$~\cite{MEG} (lower horizontal line), $B(\mu^-
  \to e^-e^-e^+) < 10^{-12}$~\cite{PDG}  (dashed line).  We also include
  constraints  from  the non-observation  of  $\mu\to  e$ conversion  in
  ${}^{48}_{22}$Ti  and   ${}^{197}_{\  79}$Au, 
  $R^{\rm Ti}_{\mu  e} <  4.3\times 10^{-12}$~\cite{Titanium}  (dash-dotted) and
  $R^{\rm     Au}_{\mu     e}     <    7\times     10^{-13}$~\cite{Gold}
  (dash-double-dotted),  as  well  as  potential limits  from  a  future
  sensitivity    to   $R^{\rm   Ti}_{\mu    e}$   at    the   $10^{-18}$
  level~\cite{PRISM}  (lower dash-dotted line). Left panel represents 
  SLFV results. In the right panel the quantum effects due to 
  $\widetilde{N}_{1,2,3}$ are ignored.}\label{f2}
\end{figure}

\begin{figure}[t]
 \centering
  \includegraphics[clip,width=0.45\textwidth,height=0.19\textheight]{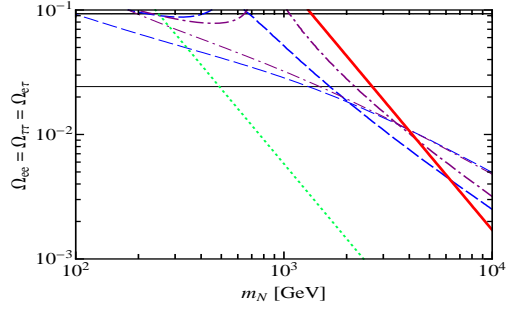}
\caption{Exclusion  contours of  $\mbox{\boldmath{$\Omega$}}_{e\tau}$ versus  $m_N$
  for $\mbox{\boldmath{$\Omega$}}_{\ell\ell'}$ values defined in
  the text. The limits are similar for a complementary
  scenario with $e$ replaced by $\mu$. SLFV limits and non-SUSY limits on
  $\mbox{\boldmath{$\Omega$}}_{\tau e}$ are represented by thicker and thiner
  lines respectively. We use the experimental upper limits~\cite{PDG}
  on  $B(\tau^- \to  e^-\gamma)  < 1.1  \times  10^{-7}$ (solid  lines),
  $B(\tau^-  \to  e^-e^-e^+)  <  3.6\times 10^{-8}$~(dashed  lines)  and
  $B(\tau^-   \to  e^-\mu^-\mu^+)   <   3.7\times  10^{-8}$~(dash-dotted
  lines).}\label{f3}
\end{figure}

We    now    present   predictions    for    the   LFV    observables:
$\ell\to\ell\gamma$~\cite{IP}, the  lepton number conserving processes
$\ell\to\ell'\ell_1\ell_2$~\cite{IP},  and the  rate  $R_{\mu e}$  for
$\mu\to  e$  conversion in  nuclei~\cite{COKFV,PRL_IP}.   We fix  $\mu
= \widetilde{M}_Q  = M_{\tilde{\nu}} =  200$~GeV, $M_{\widetilde{W}} =
100$~GeV and  $\tan\beta = 3$.  The impact of  SLFV on $\mu\to  e$ and
$\tau\to e$ is  presented in Figs.~\ref{f2} and~\ref{f3} respectively.
The  diagonal dotted lines  indicate the  regime where  terms $\propto
(\mbox{\boldmath{$\Omega$}}_{\ell\ell'})^2$     dominate    the    LFV
observables, whilst the area  above the diagonal solid lines represent
a non-perturbative  regime with ${\rm Tr}\,  ({\bf h}^\dagger_\nu {\bf
h}_\nu) >  4\pi$, which  limits the validity  of our  predictions. The
areas above or within contours are exluded.

Limits from  the absence  of $\mu\to e$  transitions are  presented in
Fig.~\ref{f2}.        We       assume       a      scenario       with
$\mbox{\boldmath{$\Omega$}}_{ee}  = \mbox{\boldmath{$\Omega$}}_{\mu e}
=               \mbox{\boldmath{$\Omega$}}_{\mu\mu}$               and
$\mbox{\boldmath{$\Omega$}}_{\tau           i}          =          0$,
$i=e,\mu,\tau$~\cite{OmegaDiag}.   Fig.~\ref{f2} shows  the  impact of
SLFV  on $\mu\to  e$ decays.   $B(\mu  \to e\gamma)$  leads to  weaker
$\mbox{\boldmath{$\Omega$}}_{\mu  e}$ constraints  than  non-SUSY case
and gives  no useful information  for low-scale SUSY  seesaw scenario.
$B(\mu \to eee)$ and
$R_{\mu  e}$ give much stronger constraints 
on $\mbox{\boldmath{$\Omega$}}_{\mu e}$ in SLFV than in non-SUSY case. 
$R_{\mu  e}$ still 
gives the best constraints for all $m_N$ values except for $m_N\sim 3$~GeV.
The projected PRISM experimental limit $R^{\rm   Ti}_{\mu   e}   \sim
10^{-18}$~\cite{PRISM} at $m_N\sim 10^8$~GeV reaches the sensitivities 
of order $\mbox{\boldmath{$\Omega$}}_{\mu e} \sim 10^{-10}$.

Limits from the non-observation of $\tau\to e$ transitions are
presented in Fig~\ref{f3}, assuming $\mbox{\boldmath{$\Omega$}}_{ee}
= \mbox{\boldmath{$\Omega$}}_{\tau e}
= \mbox{\boldmath{$\Omega$}}_{\tau\tau}$ and
$\mbox{\boldmath{$\Omega$}}_{\mu i} = 0$, $i=e,\mu,\tau$.  Given the
constraints~\cite{OmegaDiag}, a positive signal for $B(\tau^- \to
e^-e^-e^+)$ close to the present upper bound would signify that SLFV
originates from rather large Yukawa couplings and
$m_N\stackrel{>}{{}_\sim} 3$~TeV.

In summary,  we have shown  that low-mass right-handed  sneutrinos can
sizeably contribute to  observables of LFV.  Thanks to  SUSY, they can
significantly screen  the respective effect of the  heavy neutrinos on
the photonic $\mu$  and $\tau$ decays.  Hence SLFV  can be probed more
effectively  in  present  and   future  experiments  of  $\mu  \to  e$
conversion in nuclei.  The 3-body  decay observables, such as $\mu \to
eee$ and $\tau \to eee$, provide valuable complementary information on
LFV.   In particular, the  former eliminates  a kinematic  region that
remains  unprobed in  the  non-SUSY  case by  $\mu  \to e$  conversion
experiments.





\bibliographystyle{aipproc}   


\IfFileExists{\jobname.bbl}{}
 {\typeout{}
  \typeout{******************************************}
  \typeout{** Please run "bibtex \jobname" to optain}
  \typeout{** the bibliography and then re-run LaTeX}
  \typeout{** twice to fix the references!}
  \typeout{******************************************}
  \typeout{}
 }



\end{document}